 \documentclass{aa}
\usepackage{graphicx}
\begin{document}

%%%%%%%%%%%%%%%%%%%%%%%% Title Page begins %%%%%%%%%%%%%%%%%%%%%%%%%%%%%

\title{
Spatial distribution of emission in Unidentified Infrared Bands
from Midcourse Space Experiment Survey}

\author{S.K. Ghosh \and D.K. Ojha}

 \offprints{S.K. Ghosh, \email{swarna@tifr.res.in}}

\institute{Tata Institute of Fundamental Research, Homi Bhabha Road,  
Mumbai (Bombay) 400 005, India }

\date{Accepted 12 March 2002 }

\abstract{
 Recently the Midcourse Space Experiment (MSX) has surveyed
the Galactic plane in mainly four infrared bands between 6 and 25 $\mu$m.
Two of these bands cover several Unidentified Infrared 
emission Bands (UIBs).
With the aim of
extracting the spatial distribution of the UIB 
emission on a large scale,
a scheme has been developed 
to model the MSX data with emission
in the UIBs alongwith the underlying thermal continuum
from the interstellar dust.
In order to test this scheme,
a sample of five Galactic compact H II regions
(Sh-61, Sh-138, Sh-152, Sh-156, Sh-186; Zavagno \& Ducci 2001)
for which imaging study in some individual UIBs 
is available from ISOCAM measurements, has been studied.
The results of this comparative study on small angular scale
are as follows :
(i) the morphological details extracted from our scheme
 agree very well with those from the superior ISOCAM
measurements; 
(ii) the integrated strength of UIBs extracted from the
MSX database correlates extremely well with
the sum of the strengths of individual UIBs 
measured from ISOCAM.
This tight correlation is very encouraging and promises the 
potential of MSX database for study of large scale 
spatial distribution of
UIB emission (and the carriers of UIBs) in the 
entire Galactic plane.
 
}

 \titlerunning{Emission in UIBs from MSX data}
 \authorrunning{S.K. Ghosh \& D.K. Ojha}
\maketitle

%%%%%%%%%%%%%%%%%%%%%%%% Main Text begins %%%%%%%%%%%%%%%%%%%%%%%%%%%%%%
  
\section{Introduction}

 The near to mid infrared spectrum originating from the interstellar medium 
of the Galactic star forming regions 
consists of various features in addition to a continuum.
The continuum emission is attributed to the thermal emission 
from interstellar dust component
and almost all narrow emission lines have been identified 
with atomic nebular, molecular, etc  transitions originating 
from the interstellar gas component.
In addition, several broader features have been detected which 
have been identified with features due to the solid state material
of the dust (e.g. silicate absorption features at $\sim$ 10 and 
$\sim$ 18 $\mu$m etc).
There exists a class of broad emission features,
sometimes called ``Unidentified Infrared emission Bands" (UIBs;
at 3.3, 6.2, 7.7, 8.6, 11.3, 12.7 $\mu$m), identity
of whose carriers and the emission mechanisms are
still a subject of active research. 
Some of these bands are widely believed to be characteristic of
the bending and stretching modes of C=C and C--H bonds in
aromatic molecules 
(e.g. fluorescent emission of 
Polycyclic Aromatic Hydrocarbons or PAHs; Leger \& Puget 1984, 
Allamandolla et al. 1985).
However, other contenders also exist in literature
(e.g. amorphous materials with aromatic 
hydrocarbon; Sakata et al. 1984, Borghesi et al. 1987).
The study of large scale distribution of emission in the UIBs 
in the Galactic plane
in general and selected star forming regions in particular
would be 
important in understanding details of their emission mechanism.

  The recent mission Infrared Space Observatory (ISO),
in particular the imaging camera ISOCAM, has made it possible
to study selected Galactic star forming regions in the UIBs.
The ISOCAM instrument had several filters with the passbands
selected to cover these UIBs (and also neighbouring continuum) 
so that emission in these individual features can be measured
very precisely.
However, it is unreasonable to expect Galactic plane surveys
in UIB emission using the ISOCAM, since its primary objective
was to achieve best possible (nearly diffraction limited)
angular resolution in studies of individual astrophysical sources.
As a result, largest single image from ISOCAM covers $\sim$
3\arcmin $\times$ 3\arcmin.

 With the advent of the Midcourse Space Experiment (MSX), new 
possibilities have emerged. The 
SPIRIT III instrument onboard MSX spacecraft, has surveyed the entire
Galactic Plane ($|b| < 5$\degr) in four mid infrared bands 
centered around
8.3, 12.13, 14.65 and 21.34 
$\mu$m with an angular resolution $\sim$ 18\arcsec ~ (Price et al. 2001).
These four bands are referred to as $A$, $C$, $D$ and $E$ 
spectral bands of MSX respectively (the SPIRIT III instrument
also had two additional narrower bands at 4.29 and 4.35 $\mu$m, 
called $B_1$ and $B_2$ bands).
The usefulness of the MSX survey to the study of diffuse 
interstellar medium and global characteristics has already
been demonstrated (Cohen \& Green 2001; Cohen 1999).
The MSX band $A$ includes the dominant UIB features at
6.2, 7.7 and 8.7 $\mu$m. Similarly the MSX band $C$ includes 
the UIB features at 11.3 and 12.7 $\mu$m.

Whereas, the ISOCAM provides imaging capability in narrower spectral bands 
at higher angular resolution (3\arcsec ~ or 6\arcsec ~) of 
selected regions,
the MSX survey covers the entire Galactic plane in four broader
bands (two of these covering several UIBs in addition to the continuum).
 Making use of this complementarity of ISOCAM vis a vis 
MSX, the following scheme has been explored to study 
large scale emission in the UIBs in the Galaxy :

\begin{itemize}

\item Model each picture element of MSX images with an integrated
emission in UIB features superposed on a gray body
continuum spectrum under some reasonable assumptions. 
The best fit solution (for each pixel) provides a measure
of the UIB emission locally.  
\item Test the reliability of the above scheme by comparing
the results of some
selected Galactic star forming regions which have been
studied using the ISOCAM and whose emission in individual
UIBs have been quantified (e.g. Zavagno \& Ducci 2001).
\item The comparison should cover not only qualitative (e.g. structural
details / morphology) aspects, but also quantitative
correlation between the integrated estimate of
UIB emission from our scheme and the ISOCAM results.

\end{itemize}

 The section 2 describes the modelling scheme in detail, and the
section 3 presents the results for the sample of six Galactic 
star forming regions using the MSX survey data. 
In section 4, a comparison between our results have been made
with those from the ISOCAM by Zavagno and Ducci (2001).
The conclusions are summarised in section 5.

\section{The Scheme}

 The publicly available MSX Galactic plane survey 
radiance images (Infrared Processing and Analysis Center
at http://irsa.ipac.caltech.edu/applications/MSX)
in the four 
bands at 8.3 ($A$), 12.1 ($C$), 14.7 ($D$) and 21.3 ($E$) $\mu$m 
are gridded in 6\arcsec 
$\times$ 6\arcsec ~ pixel, though the true angular resolution is 
$\sim$ 18.3\arcsec ~ (Price et al. 2001), in the unit of
$Wm^{-2}sr^{-1}$.
The zodiacal background has already been subtracted out from these
MSX survey maps.
The spectrum emitted from each pixel is assumed to be a 
combination of a thermal continuum (modified Planck 
function or gray body) and the total radiance due to the relevant
UIB features within the MSX band.
$$R_{i} = r^{UIBs}_{i} + \int
(1-e^{-\tau_{\nu}}\,) \times B_{\nu}(T)\, \times 
RSR_{i}(\nu)\,d\nu  \eqno(1) $$
$$ \,\,\,\,\,\,\,\,\,\,\,\,\,\,\,\,\,\,\,\,\,\,\, i = A,\, C,\, D,\, \,E$$
where $R_i$ are the measured radiances 
in the MSX bands.
$B_{\nu}(T)$ is the Planck function, and the term in parenthesis
emulates the gray body spectrum emitted by the dust grains. 
$\tau_{\nu}$ is the optical depth due to the 
interstellar dust component at the frequency $\nu$.
The functions $RSR_{i}(\nu)$ represent the normalized relative spectral
responses of the four MSX bands (Egan et al. 1999).

Since the range of frequencies covered by the MSX bands is
limited, we assume a power law dependence of the
dust emissivity on frequency, viz.,
$$\tau_{\nu} = \tau_{10} \times {\left({\nu} 
                 \over {\nu_{10}}\right)}^{\beta}. \eqno(2) $$
here $\tau_{10}$ is the optical depth at 10 $\mu$m and 
$\nu_{10}$ the frequency corresponding to wavelength 10 $\mu$m.
The value of $\beta$ is a constant determined from the type
of dust assumed. The effect of varying $\beta$ is discussed later.
The $r^{UIBs}_{i}$ are the modelled total radiances in UIBs within 
the $i$-th MSX band.
Since there are no known UIBs within the bands $D$ and $E$ of MSX,
$r^{UIBs}_{D}$ = 0  and $r^{UIBs}_{E}$ = 0.
In addition, it has been assumed that the total radiance due to 
the UIBs in band $C$ is proportional to that in band $A$, viz.,
$$r^{UIBs}_{C} = \alpha \times r^{UIBs}_{A}. \eqno(3)$$
Here $\alpha$ is held fixed to a reasonable value (based on
available observational data and understanding) for all pixels of 
all star forming regions studied, though the effect of changing
the value of $\alpha$ has also been discussed later.

 For each pixel on the sky 
with sufficient signal to noise ratio in each of the
four MSX bands (implemented by map dynamic range cuts), 
we solve the set of four equations
(eq. 1), for the three unknown variables viz., $T$, $\tau_{10}$ and
$r^{UIBs}_{A}$. A non-linear chi-square minimization scheme 
based on the finite difference Levenberg-Marquardt algorithm 
has been used for this purpose. 
The integrals in equation 1 are evaluated numerically using 
a cautious adaptive Romberg extrapolation method.
In order to ensure that the best 
solution obtained indeed corresponds to a global minimum
of chi-square, the computations are repeated for 125 
different sets of initial guesses comprising of 5
values each of $T$, $\tau_{10}$ and $r^{UIBs}_{A}$. 
The grid of initial guess values for these three variables
have been selected to cover a wide range of physical 
situations (e.g. $T$ ranging between 50 and 800 K;
$\tau_{10}$ between $10^{-6}$ and $10^{-2}$;
$r^{UIBs}_{A}$ between $10^{-7}$ and $10^{-3}$
$Wm^{-2}sr^{-1}$).

Invariably, the same solution is obtained starting from 
almost all different sets of initial guesses.
The procedure is repeated for all pixels of the MSX map
resulting in spatial distributions of these three physical variables.
Here we extensively use the map of $r^{UIBs}_{A}$, which
can be compared with the measurements from ISOCAM.

\section{Results}
\subsection{Results from ISOCAM}
\subsubsection{Available results from the literature \& sample
 selection}

 From the literature we have selected the work of
Zavagno and Ducci (2001; hereafter ZD) which
is based on ISOCAM measurements,
for detailed comparison with results from our scheme of extracting
emission in UIBs from the MSX data.
The reason for selecting ZD is that they 
very uniformly studied (using same set of filters) a reasonable
sample size comprising of five Galactic compact H II regions,
viz., Sharpless(Sh)-61, Sh-138, Sh-152, Sh-156 and Sh-186.
They studied the entire 3--12 $\mu$m wavelength range
accessible through ISOCAM. The ZD sample of compact H II regions
are bright in IRAS 12 $\mu$m band and are known to be 
strong emitters of UIBs at 7.7, 8.6 and 11.3 $\mu$m but
show no silicate absorption feature at 10 or 18 $\mu$m.
In addition, they
represent a sequence in equivalent stellar type of the
main exciting star. 

 Results of ZD are based on imaging with 3\arcsec $\times$
3\arcsec ~ pixel mode of ISOCAM covering 87\arcsec $\times$ 87\arcsec ~ 
regions around Sh-61, Sh-152  \& Sh-186 and 174\arcsec $\times$
87\arcsec ~ around Sh-138 \& Sh-156. Based on images in 
SW1 (centre $\lambda$ : 3.57 $\mu$m; $\lambda$ range : 3.05--4.10 $\mu$m;
C\'esarsky et al. 1996), SW2 (3.30; 3.20--3.40),
LW4 (6.00; 5.50--6.50), LW6 (7.75; 7.00--8.50), 
LW8 (11.4; 10.7--12.0) filters and five selected wavelengths using the
CVF, they have quantified the UIB fluxes (actually radiances)
in the 3.3, 6.2, 7.7 and 11.2 $\mu$m features integrated over the
mapped regions.

\subsubsection{Our estimation of 7.7 $\mu$m UIB feature emission}

 Using the publicly available ISOCAM data of the ZD sample sources
(ISO Postcards from ISO Data Archive for General Users;
http://www.iso.vilspa.esa.es), we have extracted the spatial distribution of
emission in the 7.7 $\mu$m UIB feature, using a method
similar (but not identical) to ZD. First of all the
underlying continuum at 7.7 $\mu$m has been estimated from power law 
interpolation using the CVF images at 6.91 and 8.22 $\mu$m.
Next, this continuum has been subtracted from the LW6 image
and the resulting emission has been attributed to the 7.7 $\mu$m
UIB (hereafter $UIB_{7.7}$ map).
The resulting maps are presented and discussed in the next section.

%-------------------- Table 1---begins 
\begin{table*}
\begin{center}
\caption{Dynamic ranges of MSX maps and extracted peak $UIB_A$ radiance}
\vskip 0.5cm
\begin{tabular}{|c|c|c|c|c|c|c|}
\hline
Source & MSX Master & \multicolumn{4}{c|}{Usable Dynamic Range}&Peak($UIB_A$) \\
\cline{3-6}
name&Plate Number&Band $A$&Band $C$&Band $D$&Band $E$& $W.m^{-2}.Sr^{-1}$ \\

\hline

Sh-61 & GP027.0+1.5 & 41 &  20 & 25 & 40 & 7.74 $\times 10^{-5}$  \\

Sh-138 & GP105.0+0.0 & 30  & 17 & 19 & 80 & 8.35 $\times 10^{-5}$  \\

Sh-152 & GP109.5-1.5 & 52 & 26 & 27 & 39 &  5.82 $\times 10^{-5}$ \\

Sh-156 & GP109.5+0.0 & 71 & 41 & 51 & 109 & 9.93 $\times 10^{-5}$ \\

Sh-186 & GP124.5+0.0 & 15 & 7.7 & 5.7 & 8.5 & 1.70 $\times 10^{-5}$ \\

\hline
\end{tabular}
\end{center}
\end{table*}
%-------------------- Table 1---ends

\subsection{Results obtained from the MSX survey}

  The MSX images of the five sources from ZD sample were
processed following the scheme described above. 
The dust emissivity power law index $\beta$ has been 
taken to be 1.0, which is commonly used in literature 
for general interstellar grains in
the mid infrared wavelength region relevant to
MSX bands (Scoville and Kwan 1976,
Savage and Mathis 1979: Mathis 1990).
However, other values of $\beta$ have also been explored
to ensure that
the results obtained here are not sensitive to the choice
of $\beta$, as discussed later.

\subsubsection{Choice of $\alpha$}

 The value of $\alpha$ represents the ratio of radiances
in the UIBs within the passbands of MSX bands $C$ and $A$ (see
eq. 3). These correspond to the features at 11.2 and 12.7 $\mu$m
(in band $C$) and at 6.2, 7.7 and 8.6 $\mu$m (in band $A$).
The value of $\alpha$ used here is based on the recent work
by Verstraete et al. (2001) using the ISO-SWS. 
They have studied the 2.4 -- 25 $\mu$m spectra of three
selected bright Galactic interstellar regions where dense molecular
gas is illuminated by stellar radiation. Their wavelength 
range covers all the four MSX bands adequately.
In addition, their choice of the three regions, viz.,
NGC 2023, Orion bar and M17-SW spans a wide range of 
excitation parameter (flux as well as hardness of
the radiation field). The spectral resolution of the
ISO-SWS measurements used by them was either 200 or 500,
significantly large to resolve individual UIBs.
The average value of $\alpha$ for the above three
regions representing different physical conditions,
has been estimated by us from Fig.1
of Verstraete et al. (2001) to be 0.32 
(with a very small dispersion).
The effect of choosing different values of $\alpha$ have been
studied by us, which is discussed later. 

From the study of Verstraete et al. (2001)
it is clear that the contribution of forbidden ionic lines
and molecular rotational lines, to the radiance within the
MSX bands is negligible compared to the UIBs and the
underlying continuum. Hence presence of these narrow lines 
should not affect our scheme of extracting the radiance due
to the UIBs from the MSX data.

\subsubsection{Maps of extracted UIB radiance}

The extent of the angular region considered around each
target was determined by the available dynamic
ranges in the four MSX bands (i.e. only pixels satisfying
the dynamic range condition in each band, were modelled). 
The usable dynamic range, $UDR$, for each band was defined
from the frequency distribution of the
radiance values, $f(R)$, in the corresponding map in the following 
manner : 
$UDR$ = $R_{max}$ / ($R_{median}$ + $R^-_{1/2}$), where
$R_{max}$ is the brightest pixel value, $R_{median}$ 
is the median value of $R$ as determined from $f(R)$,
and $R^-_{1/2}$ represents the brightness value satisfying
$f(R^-_{1/2})$ = $f(R_{median})/2$ and 
$R^-_{1/2} < R_{median}$.
For a purly Gaussian distribution, the above translates to
using the brightest $\sim$ 15 \% pixels of the full sample.
It may be noted here that any particular choice of 
dynamic range only
changes the outer boundary of the region where our scheme is
applied by either including or excluding these pixels,
without affecting any numerical results for other pixels.
The details of the dynamic range used for the five regions
are presented in Table 1.

  The resulting integrated UIB radiance maps extracted
by us from the MSX data for 
Sh-61, Sh-138, Sh-152, Sh-156 and Sh-186 are 
displayed in Figures 1 to 5 respectively as isophot
contour plots. 
The sizes of these maps are selected based on
the regions covered by the study by ZD.
The peak UIB radiance values are also listed in Table 1.

  Morphologically the extracted UIB radiance maps are very
similar to the corresponding maps of thermal continuum emission
from the dust grains, in general (at the scale of MSX resolution).
However, there are differences between the spatial distribution
of UIB emission and the other modelled parameters, viz., the dust 
temperature and the optical depth corresponding to the thermal
continuum emission. As one example, the map of
dust optical depth ($\tau_{10}$, at 10 $\mu$m) for the Sharpless
152 region is presented in Figure 6, which can be compared 
with Figure 3.

%%%%%%%%%%%%%%%%%%%%%%% Fig1 begins
\begin {figure}
\begin {center}
%\vskip -1cm
\includegraphics[height=7.0cm]{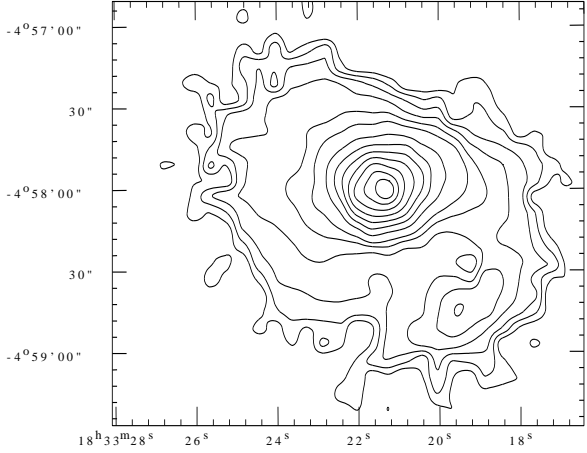}
%\vskip -0.7cm
\caption{
The spatial distribution of total radiance in Unidentified Infrared
emission Bands for the region around Sharpless 61,
as extracted from the MSX maps.
The contour levels are at 99, 90, 80, 70, 60, 50, 40,
30, 25, 20, 15, 10 \& 5 \% of the peak
value of 7.74$\times 10^{-5} W.m^{-2}.Sr^{-1}$.
The abscissa and the ordinate are R.A.(J2000.0) and 
Dec.(J2000.0) respectively.
}
\vskip -0.5cm
\end{center}
\end {figure}
%%%%%%%%%%%%%%%%%%%%%%% Fig1 Ends

%%%%%%%%%%%%%%%%%%%%%%% Fig2 begins
\begin {figure}
\begin {center}
%\vskip -1cm
\includegraphics[height=7.0cm]{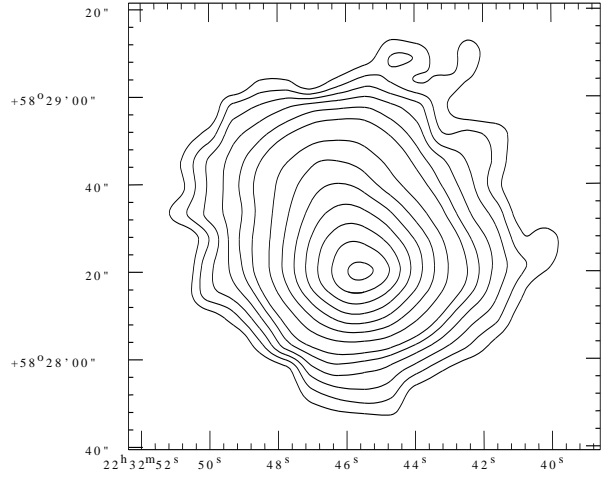}
%\vskip -0.7cm
\caption{
Same as Fig. 1 but for Sharpless 138 region.
The contour levels represent the same fractions of the 
peak as in Fig. 1. 
The peak here is 8.35$\times 10^{-5} W.m^{-2}.Sr^{-1}$.
}
\vskip -0.8cm
\end{center}
\end {figure}
%%%%%%%%%%%%%%%%%%%%%%% Fig2 Ends

%%%%%%%%%%%%%%%%%%%%%%% Fig3 begins
\begin {figure}
\begin {center}
%\vskip -1cm
\includegraphics[height=7.0cm]{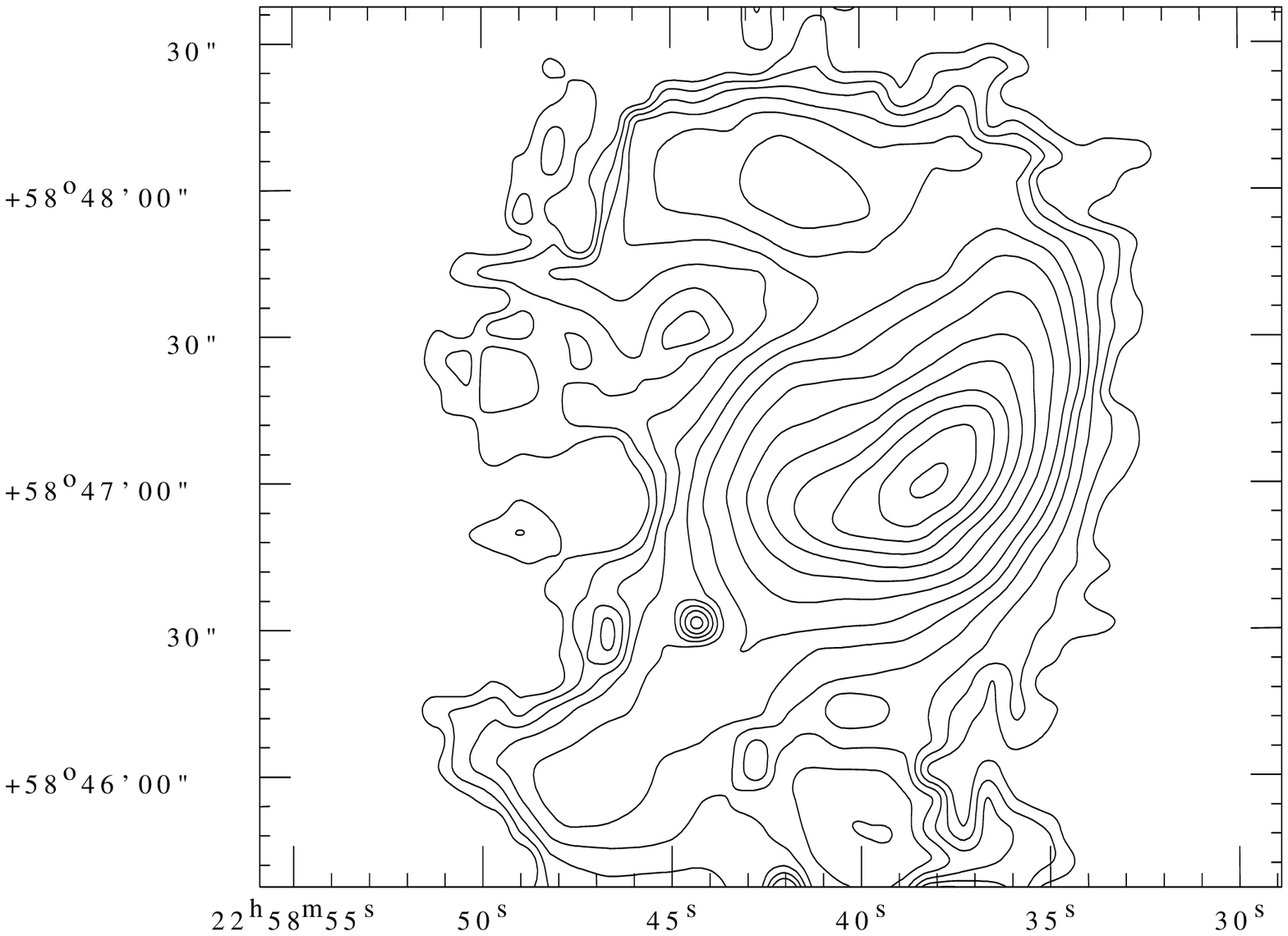}
%\vskip -0.7cm
\caption{
Same as Fig. 1 but for Sharpless 152 region.
The peak here is 5.82$\times 10^{-5} W.m^{-2}.Sr^{-1}$.
}
\vskip -0.8cm
\end{center}
\end {figure}
%%%%%%%%%%%%%%%%%%%%%%% Fig3 Ends

%%%%%%%%%%%%%%%%%%%%%%% Fig4 begins
\begin {figure}
\begin {center}
%\vskip -1cm
\includegraphics[height=7.0cm]{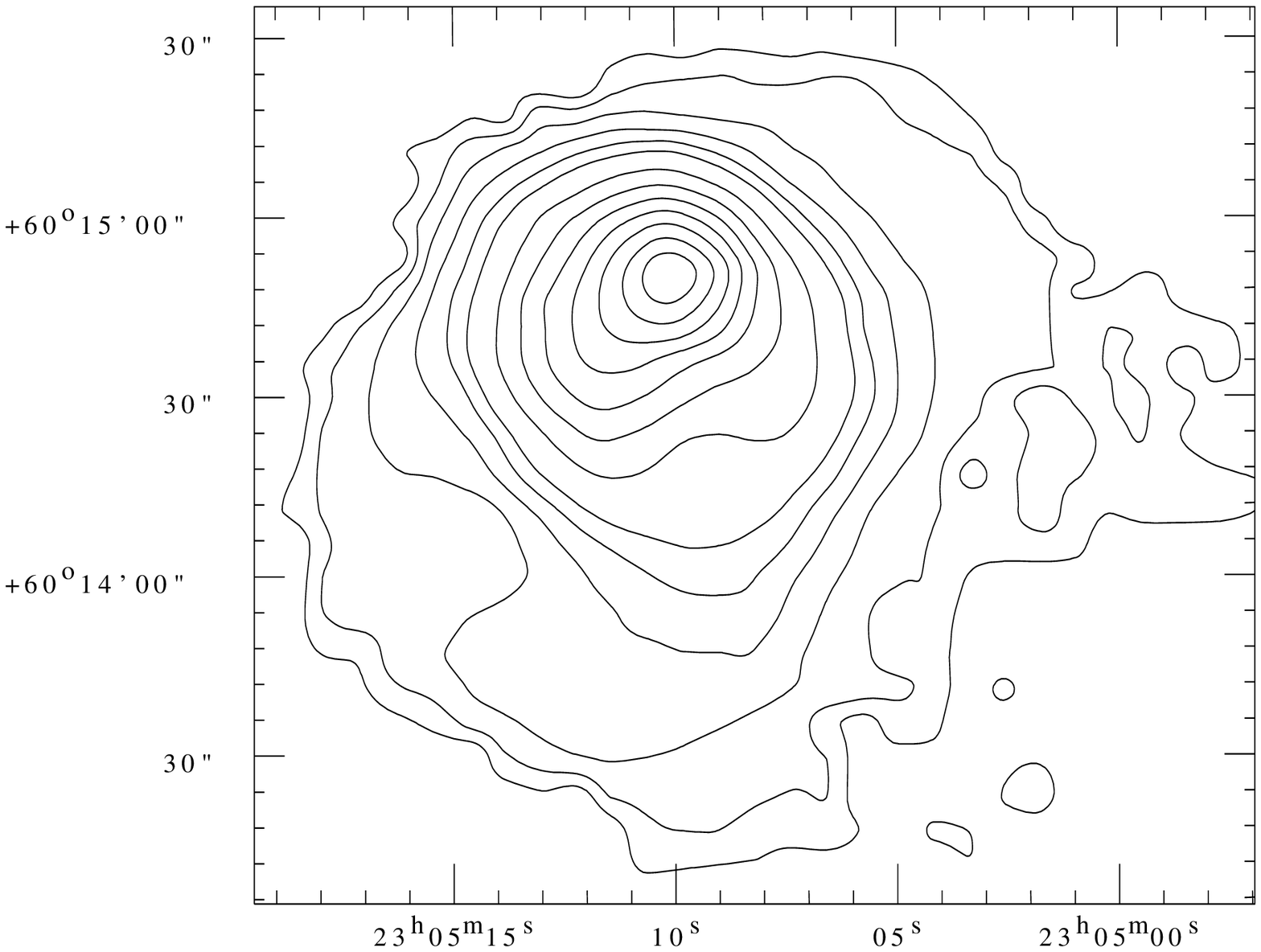}
%\vskip -0.7cm
\caption{
Same as Fig. 1 but for Sharpless 156 region.
The peak here is 9.93$\times 10^{-5} W.m^{-2}.Sr^{-1}$.
}
\vskip -0.5cm
\end{center}
\end {figure}
%%%%%%%%%%%%%%%%%%%%%%% Fig4 Ends

%%%%%%%%%%%%%%%%%%%%%%% Fig5 begins
\begin {figure}
\begin {center}
%\vskip -1cm
\includegraphics[height=7.0cm]{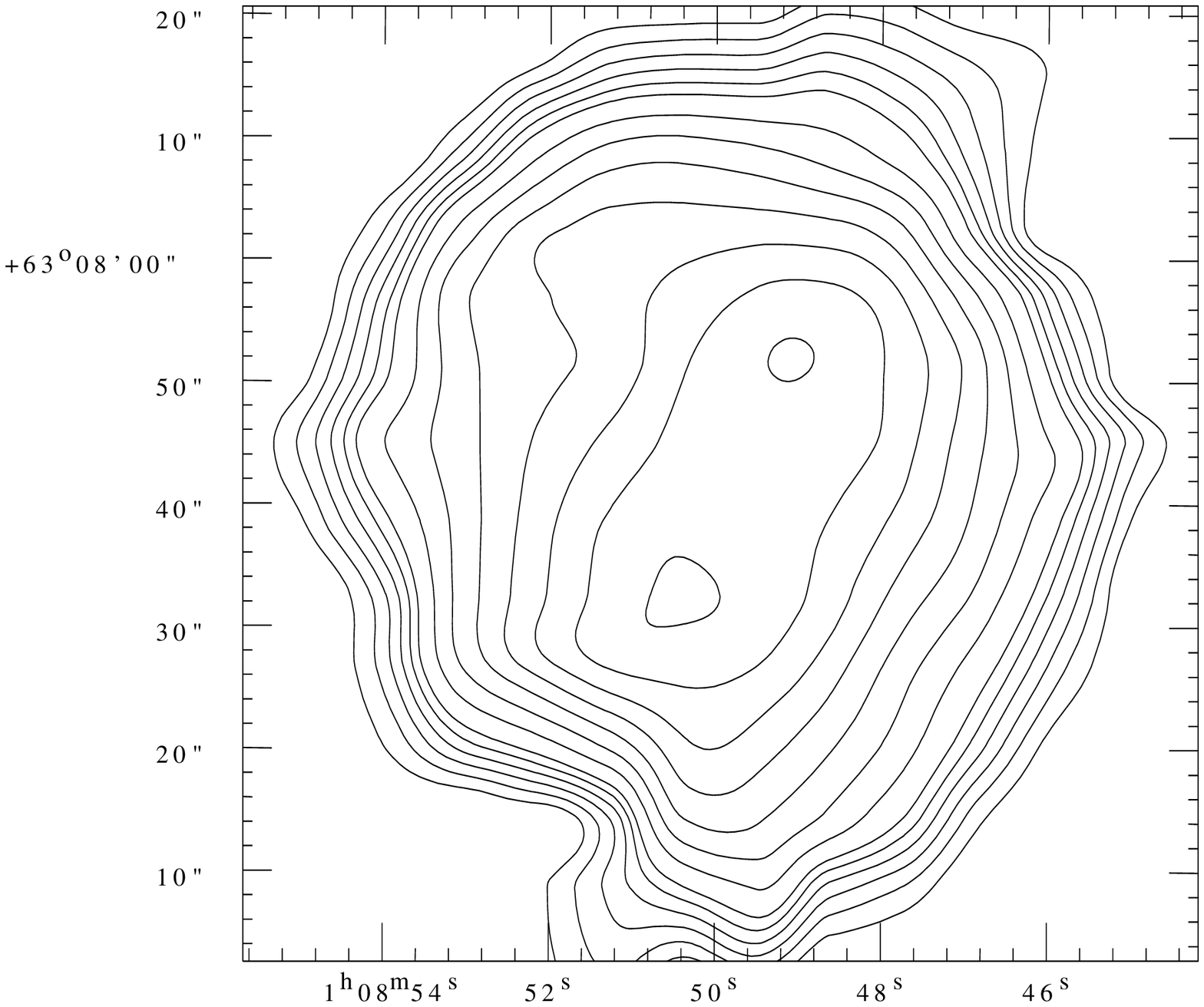}
%\vskip -0.7cm
\caption{
Same as Fig. 1 but for Sharpless 186 region.
The peak here is 1.70$\times 10^{-5} W.m^{-2}.Sr^{-1}$.
}
\vskip -0.5cm
\end{center}
\end {figure}
%%%%%%%%%%%%%%%%%%%%%%% Fig5 Ends

%%%%%%%%%%%%%%%%%%%%%%% Fig6 begins
\begin {figure}
\begin {center}
%\vskip -1cm
\includegraphics[height=7.0cm]{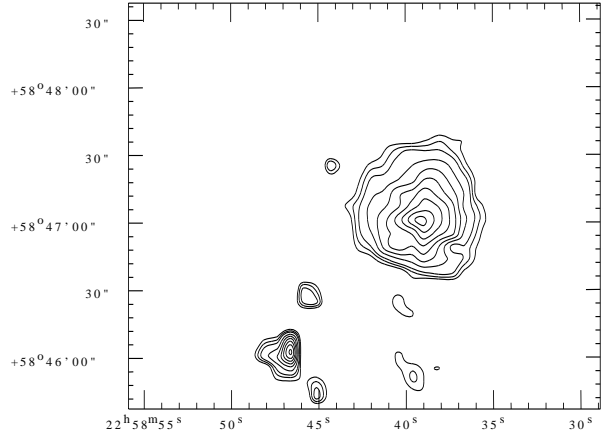}
\caption{
The spatial distribution of dust optical depth ($\tau_{10}$
at 10 $\mu$m) for the Sharpless 152 region.
The contour levels are at 99, 90, 80, 70, 60, 50,
40, 30 \& 20 \% of the peak value of 3.49$\times 10^{-4}$.
The structure of these contours differ from those 
of UIB radiance for the same region shown in
Figure 3.
}
\vskip -0.5cm
\end{center}
\end {figure}
%%%%%%%%%%%%%%%%%%%%%%% Fig6 Ends

\section{Comparison between MSX and ISOCAM results}

\subsection{Morphological similarities}

It is instructive to compare our maps from 
modelling of MSX data with those
based on the ISOCAM data.
Most relevant ISOCAM filters for this comparison
are LW4, LW6 and LW8 covering the 6.2, 7.7 and
11.2 $\mu$m UIBs. The former two features contribute 
to the UIB radiance in MSX band $A$ and the last one
in MSX band $C$.

The ISOCAM based $UIB_{7.7}$ maps have been superposed (grey scale)
on our integrated UIB map extracted from MSX maps, viz. $r^{UIBs}_{A}$,
(hereafter $UIB_A$), in Figures 7 to 11 corresponding to 
Sh-61, Sh-138, Sh-152, Sh-156 and Sh-186 respectively.
For a more general comparison involving other UIBs, we refer
to Figure 2 of ZD.
In principle, our $UIB_{7.7}$ map for each source must resemble LW6
map of ZD,
which indeed is the case.

%%%%%%%%%%%%%%%%%%%%%%% Fig7 begins
\begin {figure}
\begin {center}
%\vskip -1cm
\includegraphics[width=8.0cm]{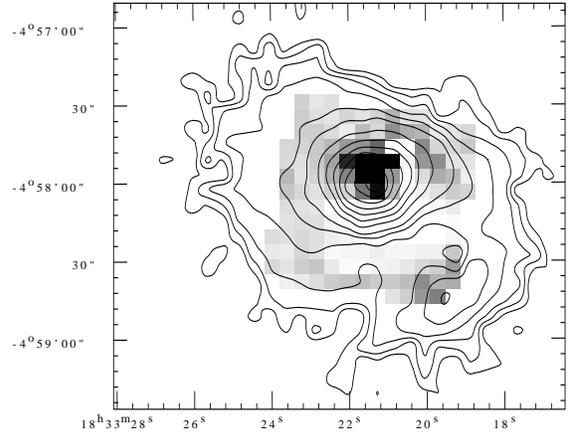}
%\vskip -0.7cm
\caption{
Comparison of the
 total radiance in Unidentified Infrared
emission Bands ($UIB_A$; contours, same as in Fig. 1) 
as extracted from the MSX maps, with
the emission in the 7.7 $\mu$m UIB feature
obtained from ISOCAM ($UIB_{7.7}$ in grey scale;
the grey scaled region also represents the area imaged by
ISOCAM),
for the region around Sharpless 61.
The abscissa and the ordinate are R.A.(J2000.0) and 
Dec.(J2000.0) respectively.
}
\vskip -0.5cm
\end{center}
\end {figure}
%%%%%%%%%%%%%%%%%%%%%%% Fig7 Ends

%%%%%%%%%%%%%%%%%%%%%%% Fig8 begins
\begin {figure}
\begin {center}
%\vskip -1cm
\includegraphics[width=8.0cm]{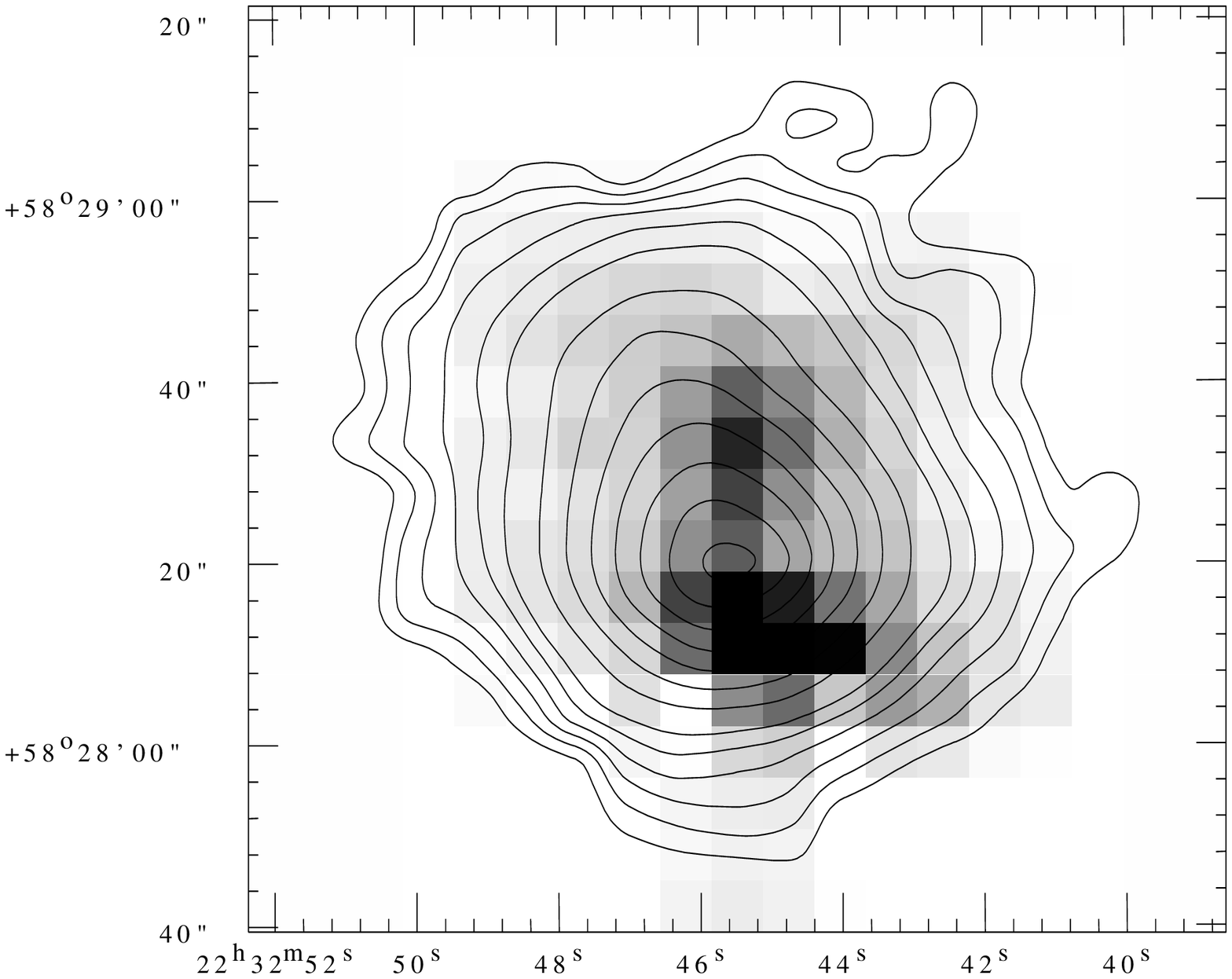}
%\vskip -0.7cm
\caption{
Same as Figure 7, but for the Sharpless 138 region.
}
\vskip -0.8cm
\end{center}
\end {figure}
%%%%%%%%%%%%%%%%%%%%%%% Fig8 Ends
%%%%%%%%%%%%%%%%%%%%%%% Fig9 begins
\begin {figure}
\begin {center}
%\vskip -1cm
\includegraphics[width=8.0cm]{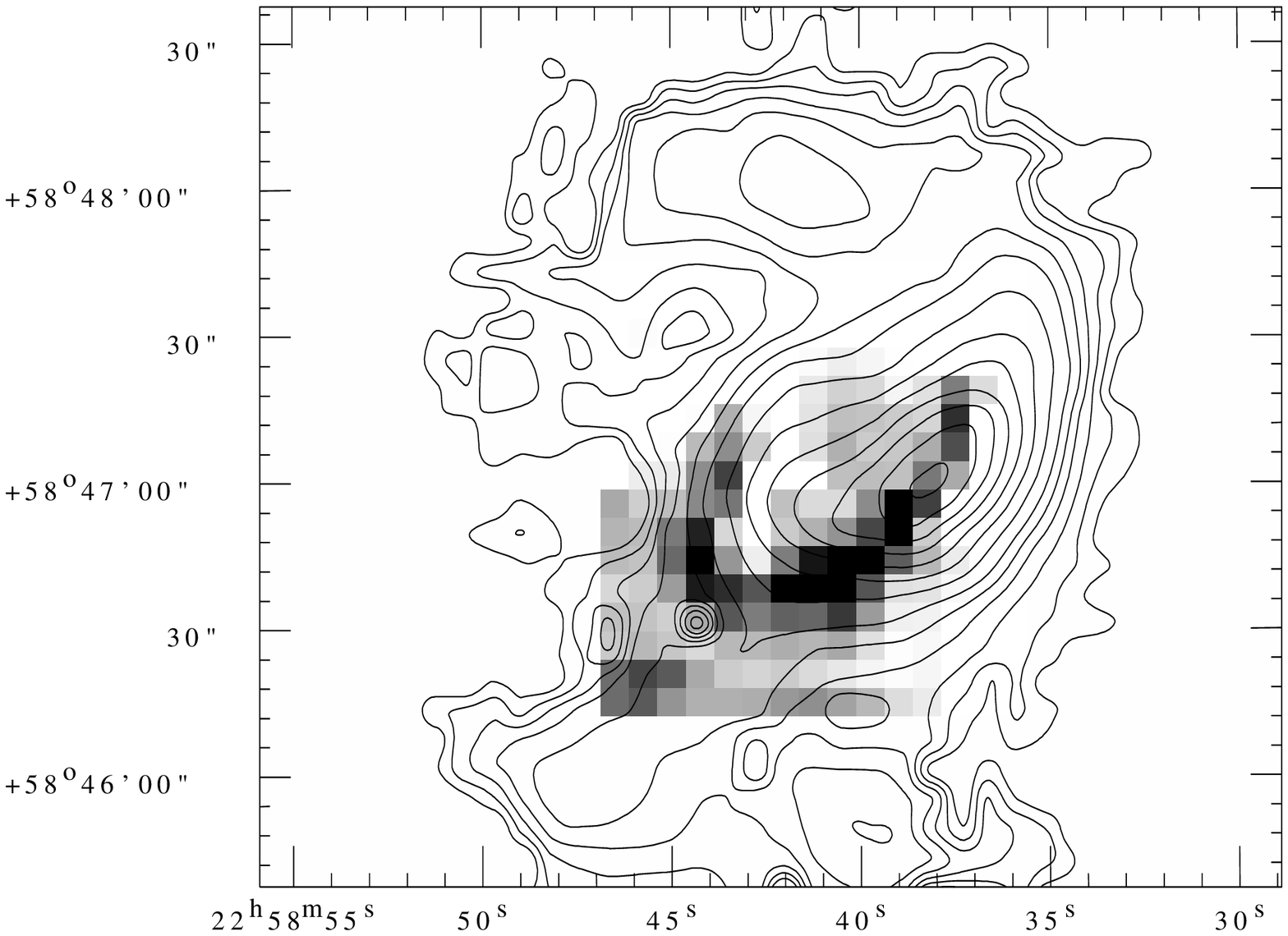}
%\vskip -0.7cm
\caption{
Same as Figure 7, but for the Sharpless 152 region.
}
\vskip -0.5cm
\end{center}
\end {figure}
%%%%%%%%%%%%%%%%%%%%%%% Fig9 Ends
%%%%%%%%%%%%%%%%%%%%%%% Fig10 begins
\begin {figure}
\begin {center}
%\vskip -1cm
\includegraphics[width=8.0cm]{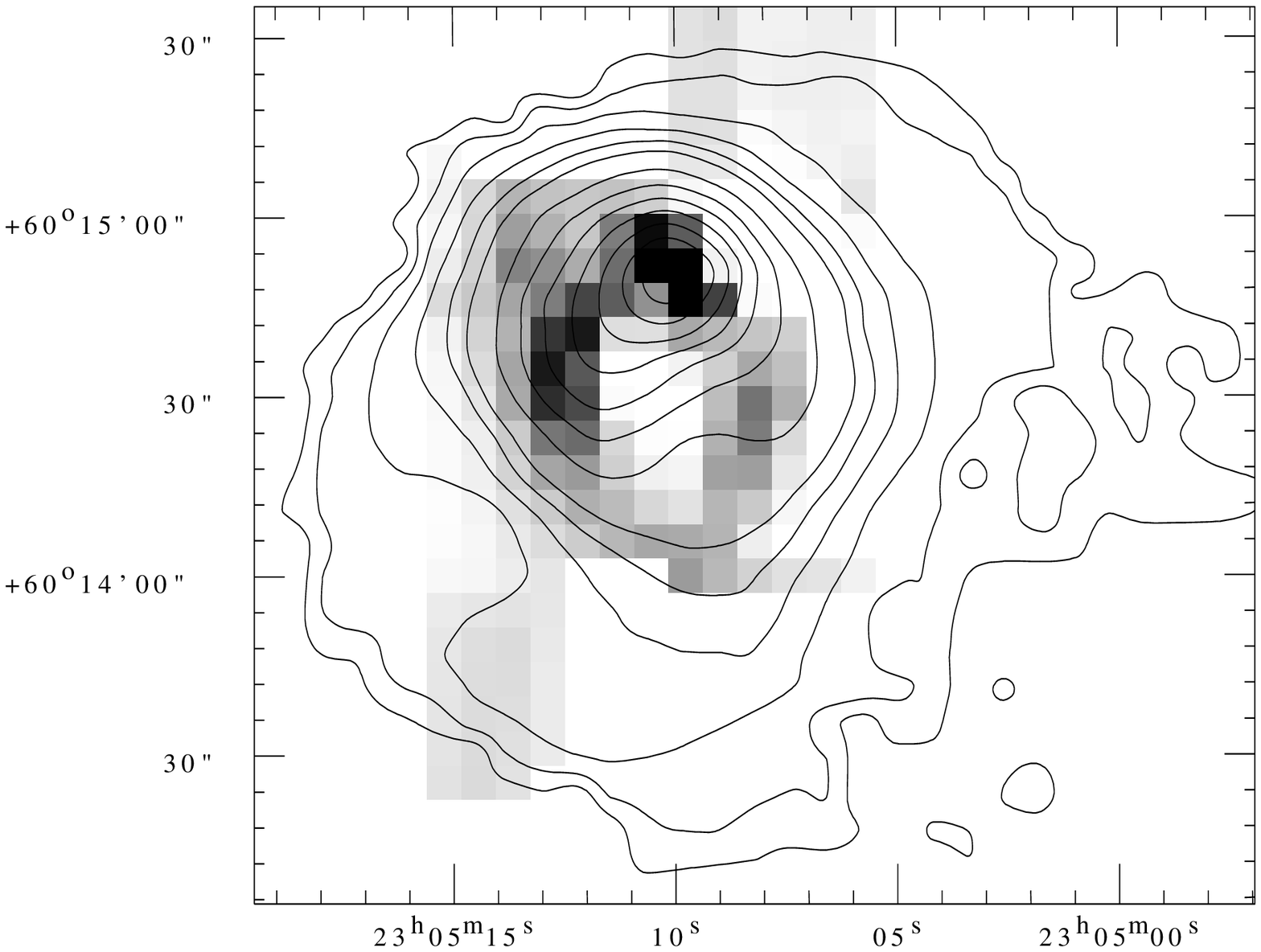}
%\vskip -0.7cm
\caption{
Same as Figure 7, but for the Sharpless 156 region.
}
\vskip -0.5cm
\end{center}
\end {figure}
%%%%%%%%%%%%%%%%%%%%%%% Fig10 Ends
%%%%%%%%%%%%%%%%%%%%%%% Fig11 begins
\begin {figure}
\begin {center}
%\vskip -1cm
\includegraphics[width=8.0cm]{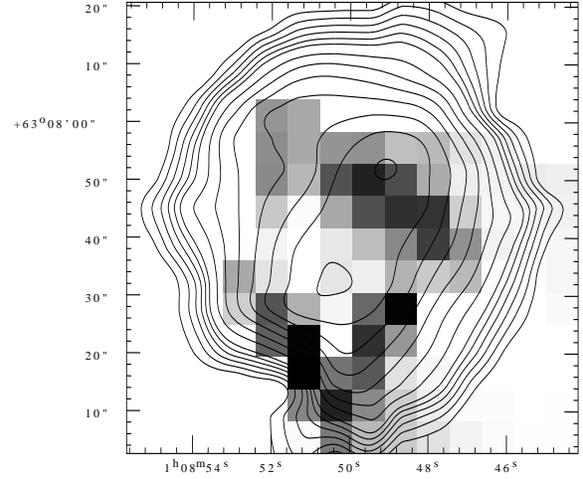}
%\vskip -0.7cm
\caption{
Same as Figure 7, but for the Sharpless 186 region.
}
\vskip -0.5cm
\end{center}
\end {figure}
%%%%%%%%%%%%%%%%%%%%%%% Fig11 Ends

A marked morphological similarity between the contour plots
corresponding to all these three ISOCAM filters (LW4, LW6 \& LW8)
for all the five sources, support our assumption 
that $r^{UIBs}_{C}$ and $r^{UIBs}_{A}$ are correlated (equation 3).
Next, we comment qualitatively about 
structural similarities between the spatial distribution
of emission in individual UIBs as obtained from ISOCAM 
(our $UIB_{7.7}$; LW4, LW6 \& LW8 maps of ZD), 
vis a vis the total emission due to all UIBs within 
the MSX band $A$ ($UIB_{A}$). It may be noted that whereas ZD's
maps represent intensity while ours are radiance.
For qualitative  comparison of structural details,
it may be acceptable, however quantitive comparison
is made in identical units later in this subsection.

\subsubsection{Sharpless 61}
The peak position for Sh-61 in $UIB_A$ map matches 
exactly with that of $UIB_{7.7}$ (Fig. 7 here) / LW6 and LW4
(Fig. 2 of ZD). General structure of the
$UIB_A$ contours is also very similar to the
maps LW4, LW6 as well as LW8 of ZD, with extension along
E-W. We have detected an additional emission region due S-W from the
main peak in the $UIB_A$ map, which is perhaps barely
below the lowest contour displayed in LW4 or LW6.
However, our $UIB_{7.7}$ map gives some hint of the same
despite being near the edge of the ISOCAM field.

\subsubsection{Sharpless 138}
For Sh-138 (Fig. 8), 
the $UIB_{7.7}$ map has been able to resolve two components 
along N-S, and
the $UIB_A$ map shows the peak position as well as the contour shape
consistent with that
(note that the former has an angular resolution
is $\sim$ 18\arcsec ~ while ISOCAM's is 3\arcsec). 
The contours in LW4, LW6 and LW8 maps are remarkably
similar with a main peak and extension due north. 

\subsubsection{Sharpless 152}
Our $UIB_{7.7}$ (Fig. 9) as well as 
 the LW4, LW6 and LW8 maps for Sh-152 show very rich 
(but again similar between these filters) structure 
at an angular scale beyond the scope of MSX's resolution.
Still, our $UIB_A$ map for Sh-152 reproduces the major structures
consistent with the above, viz., the curved shape extending
approximately along S-E to N-W direction. 
The entire $UIB_A$ map seems to be shifted by $\sim$ 10\arcsec ~ 
with respect to $UIB_{7.7}$ map.
Additional interesting structures are detected in our map 
(see Fig. 3; due north and due south with respect to the main peak)
which lie outside the region mapped by ISOCAM.

\subsubsection{Sharpless 156}
 In the case of Sh-156 also the 
$UIB_{7.7}$, LW4, LW6 and LW8 maps show
structural details at smaller angular scale which are
unresolved in our $UIB_A$ map (Fig. 10). But the curvature and
density of contours in the latter indicate a morphology
consistent with the above ISOCAM maps.
The positions of respective peaks match very
well in all these maps.
UIB Emission is detected from a larger extended region
all around the region mapped by ISOCAM, with one isolated
secondary peak due west of the main peak (Fig. 4).

\subsubsection{Sharpless 186}
The $UIB_{7.7}$ as well as the $UIB_A$ maps 
have resolved the Sh-186 region into two major 
peaks separated mainly along S-E to N-W line (Fig. 11). 
The stronger component being the S-E one in all maps.
Similar structure is seen in the 
 LW4, LW6 and LW8 maps too.
The position of the peaks and other
structural features also match very well (within $<$ 10\arcsec)
between the $UIB_A$ and all ISOCAM maps.
The $UIB_A$ map shows much more extended emission well outside
the region mapped by ZD (Fig. 5).

\subsection{Quantitative correlation}

  In order to make quantitative comparison, the $UIB_A$
radiance maps of the five regions have been integrated over 
the same corresponding regions as imaged by ZD, to get 
I(UIB$_A^{MSX}$). 
ZD have tabulated the solid angle integrated UIB fluxes F(3.3$\mu$m),
F(6.2$\mu$m), F(7.7$\mu$m) and F(11.2$\mu$m) in their Table 4.
We have compared their [F(6.2$\mu$m)+F(7.7$\mu$m)]=
I(ISOCAM$_{4+6}$) 
for each region with our integrated $UIB_A$ radiances (see Table 2).
Surprizingly, an extremely tight linear correlation
has been found between 
I(ISOCAM$_{4+6}$) and I(UIB$_A^{MSX}$)! 
It has been found that the ratio, ${\cal R}_{A}$ =
 I(UIB$_A^{MSX}$) / I(ISOCAM$_{4+6}$) = 2.29 $\pm$ 0.07
(mean value and the error on the mean).
This is despite several simplifying assumptions made in our
analysis of the MSX data.
This is indeed a remarkable finding, considering the very
complex microscopic as well as macroscopic details that 
must go into deciding the amount of emission in the UIBs.
Whereas the UIBs at 3.3, 8.6, 11.3 and 12.7 $\mu$m originate from 
vibrational modes of the aromatic C--H bond, the bands
at 6.2 and 7.7 $\mu$m arise from the aromatic C--C bonds.
There is strong evidence from the laboratory work on 
PAHs (carriers of UIBs), that the relative strengths of these
features are very sensitive to the ionization state of the 
PAHs (Allamandola, Hudgins \& Sandford 1999).
Several observational studies of the spatial distribution
of UIBs in well resolved H II regions support the same.
Joblin et al. (1996) found that the ratio of 8.6 $\mu$m to 11.3 $\mu$m
emission increases with the local far-UV (FUV) flux, which can ionize
the PAH molecules. In a more recent study, Cr\'et\'e et al. (1999) found similar
variations in the ratio of UIB features for the M 17 complex.
One possible explanation for the above correlation is as follows :
(i) the angular sizes of the sample of sources considered here
are such that the intrinsic resolution of MSX ($\sim$ 18\arcsec )
spatially averages out various excitation effects expected close to
the source of FUV radiation field; and (ii) the sample spans a 
somewhat limited range of FUV luminosity ($\sim$ 2).
It may be interesting to extend the present work to 
nearby star forming regions for which the MSX resolution is 
adequate to probe the UIB emitting regions close to the
exciting FUV source, 
and also covering a larger range of $L_{FUV}$.

In any case, our
strong empirical correlation
has many important and useful implications.
For example, spatial distribution of emission in
the UIBs for the entire Galactic plane can be studied following
our method and the MSX survey with an angular resolution $\sim$ 
20\arcsec. Of course the ISOCAM data provides the
higher angular resolution information of 
selected regions and also the very important calibration factor
above.

%-------------------- Table 2---begins 
\begin{table*}
\begin{center}
\caption{Comparison of integrated radiances in UIBs obtained from ISOCAM
measurements and those extracted by our scheme from MSX survey$^a$}
\vskip 0.5cm
\begin{tabular}{|c|c|c|c|}
\hline
Source & I(ISOCAM$_{4+6}$)$^b$ & I(UIB$^{MSX}_{A}$)$^c$ & ${\cal R}_{A}$ \\
name& $W.m^{-2}$ & $W.m^{-2}$ & \\

\hline

Sh-61 & 1.64 $\times 10^{-12}$ & 4.00 $\times 10^{-12}$ & 2.44  \\

Sh-138 & 1.50 $\times 10^{-12}$ & 3.12 $\times 10^{-12}$  & 2.08 \\

Sh-152 & 1.74 $\times 10^{-12}$ & 3.88 $\times 10^{-12}$  & 2.23 \\

Sh-156 & 2.20 $\times 10^{-12}$ & 5.56 $\times 10^{-12}$  & 2.53 \\

Sh-186 & 3.83 $\times 10^{-13}$ & 8.38 $\times 10^{-13}$ & 2.19  \\

\hline
%\vskip 0.5cm
\end{tabular}
\end{center}
\vskip 0.5cm

$^a$ Solid angle integration has been carried out over an identical region
in both cases (ISOCAM \& MSX).\\
$^b$ From Zavagno \& Ducci (2001).\\
$^c$ Extracted from the MSX Galactic Plane Survey data using the scheme
presented here.\\
%\end{tabular}
%\end{center}
\end{table*}
%-------------------- Table 2---ends

Since the band $A$ of MSX includes the UIB at 8.6 $\mu$m
in addition to the ones at 6.2 and 7.7 $\mu$m, it is natural
to obtain a value for ${\cal R}_{A}$ greater than unity.
The mean value of ${\cal R}_{A}$ so obtained can be interpreted 
in terms of the relative strength of 8.6 $\mu$m feature
vis a vis 6.2 $\mu$m + 7.7 $\mu$m features.

A similar correlation between $UIB_C$ (which is
just a scaled down value of $UIB_A$) and the 11.2 $\mu$m
 feature emission measured using ISOCAM (LW8) by ZD has been 
explored for our sample of five sources.
The ratio, ${\cal R}_C$ = I(UIB$^{MSX}_C$)/I(ISOCAM$_8$), turns out to be
 2.98 $\pm$ 0.28, which shows a much larger
dispersion. 
In view of the fact that band $C$ of MSX includes the UIB
feature at 12.7 $\mu$m also, the above is not surprizing.
The average value of ${\cal R}_C$ should provide information 
regarding relative strengths of the features at 11.2 and
12.7 $\mu$m. The larger dispersion in the values of ${\cal R}_C$
among the five regions considered here, is perhaps indicative of the
observed variability of the strength of 12.7 $\mu$m feature vis-a-vis 
other UIBs at shorter wavelengths (e.g. among the sources whose
 ISO-SWS spectra have been presented by Verstraete et al. 2001).

Next, we discuss, how sensitive is the value of
${\cal R}_{A}$ to the various assumptions made in
 analysing / modelling the MSX data.
 One of the parameters while modelling out the
continuum from the MSX radiance in band $A$, is
$\beta$, the dust emissivity index.
Let us consider a few most popular types of dust grains,
viz., Draine and Lee type (DL; Draine \& Lee 1984),
and Mathis, Mezger and Panagia (MMP; Mathis et al. 1983).
One of the most popular dust size distribution is due to 
Mathis, Rumple and Nordsieck (MRN; Mathis et al. 1977).
The MRN size distribution averaged values of 
absorption cross sections for DL and MMP type
dust, for silicate and graphite grains
have been computed earlier (Mookerjea \& Ghosh 1999;
Mookerjea et al. 1999).
Using these for
the MSX bands $A$ to $E$, we find the effective
value of $\beta$ to be in the range 0.56 to 0.89.
In any case, all calculations were repeated
for the values of $\beta$ = 0 and 0.5 (in addition to 1.0).
This had insignificant effect on the value of ${\cal R}_A$, and
also the tightness of correlation between 
I(ISOCAM$_{4+6}$) and I(UIB$_A^{MSX}$). 
It may be noted here that in case strong silicate absorption
feature at $\sim$ 9.8 $\mu$m was important, the above correlation
could have been lost. The sample of ZD had been selected
such that silicate feature is not visible in the IRAS LRS spectrum.

 Another very important parameter in our scheme is
$\alpha$ (see equation 3), which has been held at a value
of 0.32 on the basis of result of Verstraete et al. (2001).
In order to study the sensitivity of our results on the
numerical value of $\alpha$, we have repeated the calculations
for a range of its values between 0 and 0.7. We find the following,
for the value of $\alpha$ between 0.15 and 0.35, the correlation
between I(ISOCAM$_{4+6}$) and I(UIB$_A^{MSX}$) remains very
tight, though the numerical value of ${\cal R}_{A}$ changes 
slightly between 2.0 and 2.3. For values of $\alpha$ outside
this range (0.15--0.35), the correlation
becomes much poorer and also the value of ${\cal R}_{A}$ decreases
on $either$ side of this range ! All of the above can
be understood, if the total UIB feature strength in MSX bands $A$
and $C$ are really proportional.
Hence we conclude that value of ${\cal R}_{A}$ determined here is
physically meaningful and should help quantifying the
UIB emission in the Galactic plane in general.

\section{Summary}

  A scheme has been developed to extract the contribution
 of Unidentified Infrared emission Bands (UIBs) from the
 mid infrared Galactic plane survey carried out by the
 SPIRIT III instrument onboard Midcourse Space Experiment (MSX)
 satellite in four bands. The scheme models the observations
 with a combination of thermal emission (gray body) from
 interstellar dust and the UIB emission from the gas component,
 under reasonable assumptions. Thus the spatial distribution
 of emission in the UIBs with an angular resolution $\sim$
 20\arcsec\ ~ (intrinsic to MSX survey) has been extracted.

  In order to verify the reliability of this scheme, a detailed
 comparison has been made with the results obtained by
 Zavagno \& Ducci (2001) using the ISOCAM instrument 
 onboard Infrared Space Observatory (ISO), which has 
 superior spectral
 and spatial resolutions than that of the MSX survey.
 Five Galactic star forming regions, viz., Sharpless 61 (Sh-61),
 Sh-138, Sh-152, Sh-156 and Sh-186, studied by Zavagno \&
 Ducci (2001) have been used in this comparison. 

   The following results have been found :

\begin{itemize}
\item the UIB emission extracted from the MSX data is able
 to reproduce all major structural / morphological details
 detected by ISOCAM, 
 consistent with its angular resolution;
\item a very tight linear correlation has been found between
 the integrated UIB emission extracted from MSX data and that
 obtained from ISOCAM data (hence MSX based UIB radiance 
 estimates can be calibrated well);
 The above validates our assumption regarding emissivity law
 for interstellar dust grains and is self consistent with
 the assumed proportionality of UIB emission between the bands
 $A$ and $C$ of MSX.
 The correlation may be a result of spatial averaging over 
 regions with different excitation details.
\item  the numerical value of the ratio of UIB radiances obtained 
 from MSX and ISOCAM is an important information for understanding
  relative strengths of UIBs.
\end{itemize}

 Hence, we conclude that our empirical scheme along with the MSX Galactic
 plane survey can be a powerful tool to study large scale 
 spatial distribution of UIB carriers.

\begin{acknowledgements}

 It is a pleasure to thank the referee Dr. M. P. Egan whose 
suggestions have improved the scientific content of this paper.
This research made use of data products from the Midcourse Space 
Experiment.  Processing of the data was funded by the Ballistic 
Missile Defense Organization with additional support from NASA 
Office of Space Science.  This research has also made use of the 
NASA/ IPAC Infrared Science Archive, which is operated by the 
Jet Propulsion Laboratory, California Institute of Technology, 
under contract with the National Aeronautics and Space 
Administration.

The present work is
based on observations with ISO, an ESA project with instruments funded by
ESA Member States (especially the PI countries: France, Germany, the
Netherlands and the United Kingdom) and with the participation of ISAS and
NASA.

\end{acknowledgements}

%%%%%%%%%%%%%%%%%%%%%%%%%%%% End of Main Text %%%%%%%%%%%%%%%%%%%%%%%%%%%%%%

%%%%%%%%%%%%%%%%%%%%%%%%%%%% Bibliography begins %%%%%%%%%%%%%%%%%%%%%%%%%%%

%%%%%%%%%%%%%%%%%%%%%%%%%%%% Bibliography ends %%%%%%%%%%%%%%%%%%%%%%%%%%%%%

\end{document}